# COIN TOSS MODELING

RĂZVAN C. STEFAN, TIBERIUS O. CHECHE

University of Bucharest, Faculty of Physics, P.O.Box MG-11, RO-077125,
Bucharest – Măgurele, Romania
E-mails: stefan.constantin61@gmail.com; tiberiuscheche@yahoo.com

*Abstract*. The torque-free motion of rigid body in gravitational field is analyzed. The coin lands on a soft surface (such as the palm of the hand) that allows no bouncing. The model assumes the coin as a rigid body with constant angular momentum vector in the body frame, and no air resistance. Unbiased probabilities of heads and tails are obtained for various initial conditions accessible to a usual person.

## 1. INTRODUCTION

Coin tossing has been around for as long as coins existed. In Roman times, it was known as 'navia aut caput' ('ship or head'), as many coins had a ship on a side and the head of an Roman emperor on the other side. Throughout history, coin tossing had been used as a way to decide between two arbitrary options, as it is assumed that it provides even odds to the both sides involved. A remarkable case of coin tossing being used as a way to determine the outcome of an important event is the semi-final game of the 1968 European Soccer Championship, when Italy and the Soviet Union finished the game at 0–0, after extra time. A coin was tossed to determine the finalist. Italy won and went on to become the European champion. Coin tossing is regarded as being a *random* phenomenon. It is considered that one can not predict the outcome of the coin toss. However, the tossing of a coin is inherently a deterministic process, obeying the physical laws. It is therefore natural to ask ourselves how can we determine the outcome knowing the initial state of the system. In this didactical paper, we develop a model to simulate the probabilistic coin toss. We start by solving the deterministic motion of a coin assimilated with a uniform cylindrical rigid body. Similar to the most recent studies of the coin toss problem [1–3], the dynamics is obtained by solving the torque-free rigid body motion in gravitational field. Our study takes into account the cases for which the angular momentum with respect to the coin is constant. Differently, from the treatment in Ref. [3], the orientation of the coin during the flight in gravitational field is obtained by computing the dihedral angle between the initial and at a moment coin plane. We find this way more comfortable as the involved algebra (based on Euler's rotation matrix) is less complex. Regarding simulation of the random process, we consider as source of uncertainty the range of the initial conditions that can be accessed by tossing the coin.

The paper is structured as follows. In section 2, the theory of the deterministic torque-free coin motion in gravitational field is presented. In section 3, we introduce the models for computing the random toss and discuss the simulation results for some natural initial conditions. In section 4, we present conclusions.

## 2. THE PHYSICAL SETUP

We begin by stating our main assumptions. Firstly, we assume that the equations of motion are Newton's equations, with no external source of influence, such as fluctuations of air, thermodynamic or quantum fluctuations of the coin etc. We consider that the coin lands on a soft surface that allows no bouncing, such as the palm of the hand. Air resistance is neglected. The coin is modeled as a thin *uniform* cylinder rigid body. We consider only two possible outcomes: heads or tails. As the coin is considered to be thin, our model does not allow it to fall and remain stable on its edge. We are interested in the outcomes of real tosses, so we analyze the toss under initial conditions accessible to a normal person. We will further discuss these assumptions in section 3.

To derive equation motions it is convenient to use three reference frames: the laboratory frame (LF), $XYZ$, the body frame (BF), $xyz$, and a transport frame (TF), $x_1y_2x_3$ (Fig. 1).

The BF is the principal axes frame rigidly attached to the coin. The TF is passing through the center of mass (CM), and moves by translation with respect to the LF. In the BF the moment of inertia tensor has the diagonal form:

$$\mathbf{I} = \begin{pmatrix} I_1 & 0 & 0 \\ 0 & I_2 & 0 \\ 0 & 0 & I_3 \end{pmatrix}$$

The components of the moment of inertia tensor with respect to the principal axes are as follows:

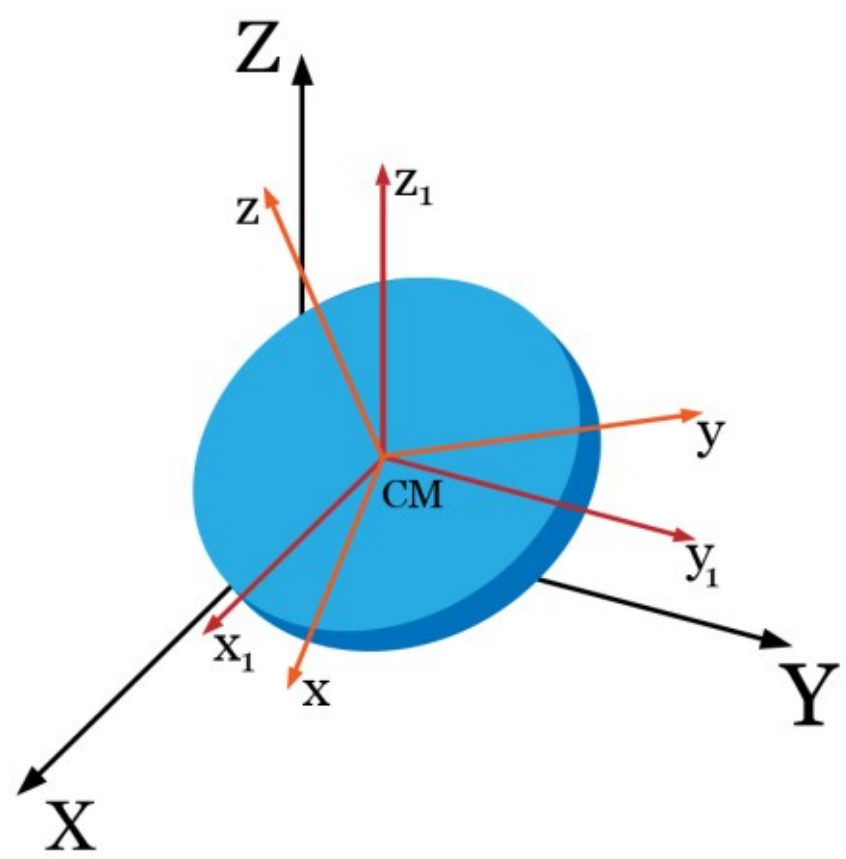

Fig. 1 – LF, TF, and BF are represented.

$$I_1 = I_2 = I_{xx} = I_{yy} = \frac{m}{4}\left(r^2 + \frac{h^2}{3}\right) \text{ and } I_3 = I_{zz} = \frac{mr^2}{2},$$

where $R$ is the radius, $h$ is the thickness, and $m$ is the mass of the cylinder.

Initially, the coin rests in equilibrium with respect to LF. Then, the coin is launched with an initial rotation. The launching is imagined as a short process, when a percussion force **F** and torques are applied to the coin. The launching process is described by using the conservation of the linear and angular momentum, **S**, with respect to the BF (which for this short duration is considered to be at rest and coincides with the TF and the LF). Thus:

$$\int_0^{\tau} \mathbf{F} \mathrm{d}t = m\Delta \mathbf{v}_c = m\mathbf{v}_{c0}, \tag{1}$$

where $\mathbf{v}_{c0}$ is the velocity of CM after percussion (the gravitational force is neglected as being much weaker than the percussion and the forces of the torques cancel each other).

In what follows, we consider a vertical percussion and consequently a vertical motion of CM, with the initial speed $v_0$ (equal to $v_{C0z}$). Generally, for the angular momentum with respect to CM, we have:

$$\frac{\mathrm{d}\mathbf{S}}{\mathrm{d}t} = \sum_j \mathbf{r}_j \times \mathbf{F}_j, \tag{2}$$

where $\mathbf{F}_j$ are all the forces applied to the body, and $\mathbf{r}_j$ are the position vectors (with respect to CM, the origin of BF) of the points where the forces are applied. If the axes passing through CM are principal (the BF axes), then

$$\mathbf{S} = \sum_{i=x,y,z} I_{ii}\omega_i \mathbf{e}_i \,, \tag{3}$$

where $\mathbf{e}_i$ are the BF unit vectors. By integration of Eq. (2), for the short time of percussion, we can write,

$$\omega_{i0} = \frac{1}{I_{ii}} \int_0^{\tau} \left[ \sum_j \mathbf{r}_j \times \mathbf{F}_j \right]_i \mathrm{d}t \,, \tag{4}$$

which explains the appearance of the initial angular velocities, $\omega_{i0}$, of the body. For didactical reasons, we emphasize that Eq. (4) is valid for the short time of percussion, when BF and TF coincides and are rest with respect to LF ( $\dot{\mathbf{e}}_i = 0$ for this case). Concluding the analysis of the initiating process, we are left with the body having a vertical initial velocity of CM and, generally, angular velocities along each axis. As the gravitational field gives zero torque in the BF, according to Euler's equations in the BF, the motion of the body continues with a torque-free motion in gravitational field.

CM of the body (or TF) has a vertical motion in gravitational field according to the law:

$$v_z = v_0 - gt \,,$$

where $g > 0$ is the gravitational attraction (vertical axis has opposite direction to gravity).

To describe the torque-free motion of the body, we consider Euler's equations with respect to BF:

$$\begin{aligned} I_1\dot{\omega}_x + \left(I_3 - I_2\right)\omega_y\omega_z &= 0 \\ I_2\dot{\omega}_y + \left(I_1 - I_3\right)\omega_x\omega_z &= 0 \,. \\ I_3\dot{\omega}_z + \left(I_2 - I_1\right)\omega_x\omega_y &= 0 \end{aligned} \tag{5}$$

For didactical reasons we emphasize that Eqs. (5) can be obtained from Eq. (2): the time derivative of **S** (written in BF, the principal axes frame) is on the left hand side and the torques of gravitational forces (which with respect to CM cancel on the total) are on the right hand side of the equations. From the symmetry of the body, Euler's equations in BF are re-written as follows,

$$\begin{aligned} I_1\dot{\omega}_x + \left(I_3 - I_1\right)\omega_y\omega_z &= 0 \\ I_1\dot{\omega}_y + \left(I_1 - I_3\right)\omega_x\omega_z &= 0 \,. \\ I_3\dot{\omega}_z &= 0 \end{aligned} \tag{6}$$

The last equation for Eqs. (6) implies that $\omega_3 = \omega_{z0}$ is constant. Eqs. (6) can easily be integrated. Thus, by a time derivative and a substitution one gets:

$$\begin{aligned} \ddot{\omega}_x + q^2 \omega_x &= 0 \\ \ddot{\omega}_y + q^2 \omega_y &= 0 \end{aligned}, \tag{7}$$

where $q = (I_3 - I_1)\omega_{z0}/I_1$. The solutions of Eqs. (7) are,

$$\omega_x = \Omega \sin(qt + \delta),\ \omega_y = \Omega \cos(qt + \delta), \tag{8}$$

and explicitly, from the initial conditions,

$$\Omega = \sqrt{\omega_{x0}^2 + \omega_{y0}^2},\ \tan\delta = \frac{\omega_{x0}}{\omega_{y0}}. \tag{9}$$

With simple algebra, from Eqs. (6), one obtains that $\boldsymbol{\omega} \cdot \mathbf{S}$ is constant. Next, by using solutions from Eq. (8), we describe particular cases of motion for constant **S**.

### 2.1. CASE a): $\omega_{z0} = 0$ AND $\omega_{y0} = 0$

In this case, as $q = 0$, from Eqs. (9) one obtains, $\Omega = \omega_{x0}$ and $\delta = \pi/2$. Then, from Eqs (8),

$$\omega_x = \omega_{x0},\ \omega_y = 0. \tag{10}$$

Thus, the BF is rotating with constant angular velocity $\omega_{x0}$ with respect to the TF or to the LF and the TF (or CM) has vertical motion in gravitational field. That is, the coin rotates about a horizontal axis that lies along *x* diameter of the coin.

To measure the rotation during the flight, we assume that the coin is caught by hand at the same height from which it was launched. Then the total duration of motion is $T = 2v_0/g$ and the total rotation angle after time interval *T* is

$$\theta(T) = T\omega_{x0} = \frac{2v_0}{g}\omega_{x0}. \tag{11}$$

Considering that the coin starts from the horizontal position with head up, we have:

i) if $\theta \in (0, \pi/2) \cup (3\pi/2, 2\pi)$ <=> $\cos\theta > 0$, then the head is up;

ii)ii) if $\theta \in (\pi/2, 3\pi/2)$ <=> $\cos\theta < 0$, then the tail is up.

### 2.2. CASE b): $\omega_{z0} = 0$

In this case, as $q = 0$, and from Eqs. (9) one obtains, $\omega_x = \omega_{x0}$, $\omega_y = \omega_{y0}$. Then, the BF is rotating with respect to the TF with constant angular velocity $\boldsymbol{\omega} = \omega_{x0}\mathbf{e}_1 + \omega_{y0}\mathbf{e}_2$. That is, the coin rotates about its diameter parallel to $\boldsymbol{\omega}$. To describe the motion with respect to the TF, we introduce the Euler angles by definition [4],

$$\begin{aligned} \omega_{x0} &= \dot{\beta}\cos\gamma + \dot{\alpha}\sin\beta\sin\gamma \\ \omega_{y0} &= -\dot{\beta}\sin\gamma + \dot{\alpha}\sin\beta\cos\gamma, \\ 0 &= \dot{\gamma} + \dot{\alpha}\cos\beta \end{aligned} \tag{12}$$

where $\alpha$ is the precession angle, $\beta$ is the nutation angle, and $\gamma$ is the spin angle. The directions of the Euler frame axes coincide with the directions of the BF axes and the Euler rotations are with respect to the TF axes. The solutions of the system of differential equations (12) depend on the initial conditions. By simple inspection of the last equation from (12), one observes that a possible solution is of the form $\alpha = \alpha_0 = \text{ct.}$ and $\gamma = \gamma_0 = \text{ct.}$ By introducing this type of solutions in Eqs. (12), one obtains that $\beta = \beta_0 + \Omega t$ with $\beta_0 = ct$ and $\gamma_0 = -\arctan(\omega_{y0}/\omega_{x0})$. To obtain position of the coin at the end of motion, we need to obtain again the dihedral angle between the coin and the horizontal plan, that we denote by $\theta$. We define the horizontal plane (with respect to the TF) by the vectors, $\mathbf{r}_A(t=0) = \mathbf{r}_{A0} = (R,0,0)$ and $\mathbf{r}_B(t=0) = \mathbf{r}_{B0} = (0,R,0)$. These vectors are rotated to $\mathbf{r}_A(t) = \mathbf{r}_A = \mathbf{A}\mathbf{r}_{A0}$ and $\mathbf{r}_B(t) = \mathbf{r}_B = \mathbf{A}\mathbf{r}_{B0}$, where $\mathbf{A}$ is the Euler rotation matrix [4]:

$$\mathbf{A} = \begin{pmatrix} \cos\gamma\cos\alpha - \cos\beta\sin\alpha\sin\gamma & \cos\gamma\sin\alpha + \cos\beta\cos\alpha\sin\gamma & \sin\beta\sin\gamma \\ -\sin\gamma\cos\alpha - \cos\beta\sin\alpha\cos\gamma & -\sin\gamma\sin\alpha + \cos\beta\cos\alpha\cos\gamma & \sin\beta\cos\gamma \\ \sin\beta\sin\alpha & -\sin\beta\cos\alpha & \cos\beta \end{pmatrix},$$

where $\alpha = \alpha_0 = 0$, $\beta = \Omega t$ and $\gamma_0 = -\arctan(\omega_{y0}/\omega_{x0})$. The dihedral angle is obtained by the scalar product,

$$\cos\theta = \mathbf{n}(\mathbf{r}_{A0}, \mathbf{r}_{B0}) \cdot \mathbf{n}(\mathbf{r}_A, \mathbf{r}_B), \tag{13}$$

where $\mathbf{n}(\mathbf{r}_{A0},\mathbf{r}_{B0}) = \mathbf{r}_{A0}\times\mathbf{r}_{B0} \big/ \left|\mathbf{r}_{A0}\times\mathbf{r}_{B0}\right|$ is the normal on the plan of vectors $\mathbf{r}_{A0}$ and $\mathbf{r}_{B0}$ (similarly for $\mathbf{n}(\mathbf{r}_A,\mathbf{r}_B)$) and the symbol '$\times$' stands for the vector product. The computation shows a simple result *independent* of angles $\alpha_0$, $\beta_0$ (that is, independent on the initial position of the coin), namely,

$$\cos\theta = \cos(\Omega t)\,. \tag{14}$$

The total rotation angle is

$$\theta(T) = T\Omega == \frac{2v_0}{g}\sqrt{\omega_{x0}^2 + \omega_{y0}^2}\,. \tag{15}$$

In addition, Diaconis' theorem [2] – according to which the angle between the normal to the coin, **n**, and angular momentum **S** remains constant – is verified. Thus, in our case, $\mathbf{S} = I_1(\omega_{x0}\mathbf{e}_1 + \omega_{y0}\mathbf{e}_2)$ and the product $\mathbf{S}\cdot\mathbf{n}$ at different times (different values of the angle $\theta$) is constant, namely, zero. One observes that the results for the case a) can be obtained as results of case b) for the particular value $\omega_{y0} = 0$.

## 3. RESULTS AND DISCUSSIONS

As we stated, we are interested in the outcome of real tosses. Within our deterministic model, we design a simulation by which we predict the probability of obtaining a certain outcome (one of the two faces of the coin) in a trial. The simulation uses the fact that, as equations (11) and (14) show, the outcome of the toss depends on two parameters: the velocity of the CM (or equivalently, duration of motion) and the angular velocity. The range of the parameter values is discretized to form configurations. In each such a configuration, the coin has a deterministic output. The output probability is computed by two ways as follows. In the first method, we used a fine discretization and computed probability as the ratio between the number of favorable (for example, heads up) configurations and total number of configurations of the discretized range (method A). For the second method, we have used a larger step for discretization but allowed a larger number of events in the trial (method B).

In order to cover a real toss, we take the speed of translation of the CM of a coin tossed in the interval of 3m/s to 5m/s (which approximately corresponds to duration of motion in the interval 0.6s to 1s). We consider that the percussion force is parallel to the vertical axis for a coin at rest and having the surface perpendicular to the vertical axis. This percussion determines a rotation about $x_1$ axis, $\omega_{x0}$, that we appreciate to be in the interval of $20\pi$ rad/s to $40\pi$ rad/s. On another hand, rotation about $y_1$, $\omega_{y0}$, is practically more difficult to be obtained by a usual person and it is expected to be approximately one order of magnitude less than $\omega_{x0}$. Next, we proceed with a number of observations about our model and the predictions it implies. For conciseness, we assume that the coin is always tossed starting head up.

In Fig. 2 we show the probability evolution with the number of configurations and coin orientations for the case a). As one can see, the probability tends to the unbiased value of $50\%$ with the increase in number of configurations by method A or in the number of events by method B. Method A, as expected from the mathematical principles, is faster convergent. We used for discretization a number of 5000 intervals for method A (total number of configurations is $5000^2$) and 100 intervals for method B.

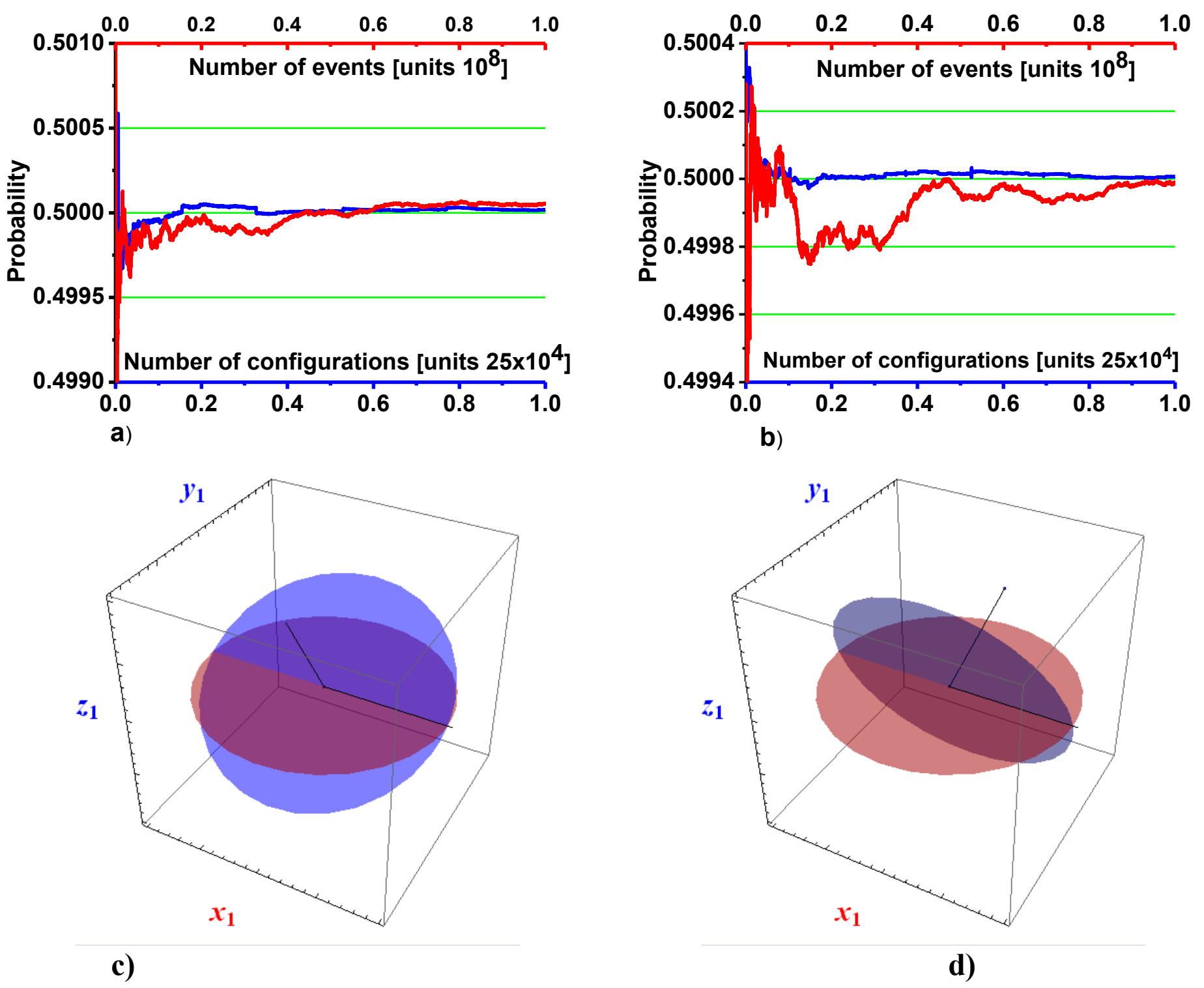

Fig. 2 – a, b – Probability computed by method A (blue) and method B (red) for the case a) ($\omega_{z0} = 0$ and $\omega_{y0} = 0$); c, d – initial (red) and at time $T$ (blue) positions of the coin. For the parameter values, see the text. (Color online)

The range of the parameters is as follows. In Fig. 2a, $\omega_{x0} \in [60, 90]$rad/s and $T = v_0/g \in [0.6, 0.9]$s. In Fig. 2b $\omega_{x0} \in [90, 120]$rad/s and $T = v_0/g \in [0.6, 0.9]$s. In Figs. 2c, d, the initial (red) and after time $T$ (blue) positions of the coin are shown for $\omega_{x0} = 80$rad/s and $T = 0.7$s, and respectively for $\omega_{x0} = 120$rad/s and $T = 0.9$s. The number of rotations in the time interval $T$ is 8.913 and 17.189, respectively.

Position of the coin is obtained with a parametric draw of the circle in the BF:

$$C(t) = R\cos(t)\mathbf{u} + R\sin(t)\mathbf{n} \times \mathbf{u},$$

where **n** is the normal vector to the circle surface (for example $\mathbf{n}(\mathbf{r}_A, \mathbf{r}_B)$) and **u** is any vector perpendicular to **n** (for example $\mathbf{r}_A$). In Fig. 2c the coin ends with head down and in Fig. 2d the coin ends with head up. As we already mentioned the rotation of the coin is about the $x_1$ axis and this is shown in Figs. 2c and 2d.

In Fig. 3 we show the probability evolution with the number of configurations and coin orientations for the case b). As in the case a), the probability also tends to the unbiased value of $50\%$ with the increase in number of configurations by method A or in the number of events by method B. The same discretization as that in case a) is used. The range of the parameters is as follows. For Fig. 3a, $\omega_{x0} \in [60, 90]$rad/s, $\omega_{y0} = 10$rad/s and $T = v_0/g \in [0.6, 0.9]$s.

For Fig. 3b, $\omega_{x0} \in [60, 90]$rad/s, $T = v_0/g \in [0.6, 0.9]$s, and to get more physical insight, we take the less probable (practical) value of $\omega_{y0} = 60$rad/s. In Figs. 3c, d, e, f the initial (red) and at time $T$ (blue) positions of the coin are shown for $\omega_{x0} = 90$rad/s, $\omega_{y0} = 30$rad/s and the durations $T = 0.6$s, $T = 0.65$s, $T = 0.7$s, and $T = 0.75$s, respectively; the number of rotations in the time interval $T$ is 9.059 and 9.8142, 10.569, and 11.324, respectively. The angular momentum, **S**, as we mentioned above, is constant (fixed direction in plane $x_1y_1$) and is perpendicular to the normal to the coin surface **n**. The directions of **S** and **n** at different times $T$ are shown in Figs. 3c, d, e, f to suggest that we have also successfully checked Diaconis' theorem.

In this work, we considered a frictionless motion. Thus, though in reality the air resistance damps the rotation and curves the trajectory of the CM, when the distance of the free fall is small the effect of the air resistance can be neglected [3]. Since our numerical values estimates height of motion in gravitational field is less than 1m, our predictions are reliable and become more accurate for smaller size coins. Another tacit assumption in calculations was that the coin can not land on its edge. As shown in a study conducted by Murray and Teare [5], it is found that a coin lands on its edge once in 6000 tosses. This low probability of the edge outcome we believe can not substantially influence our predictions.

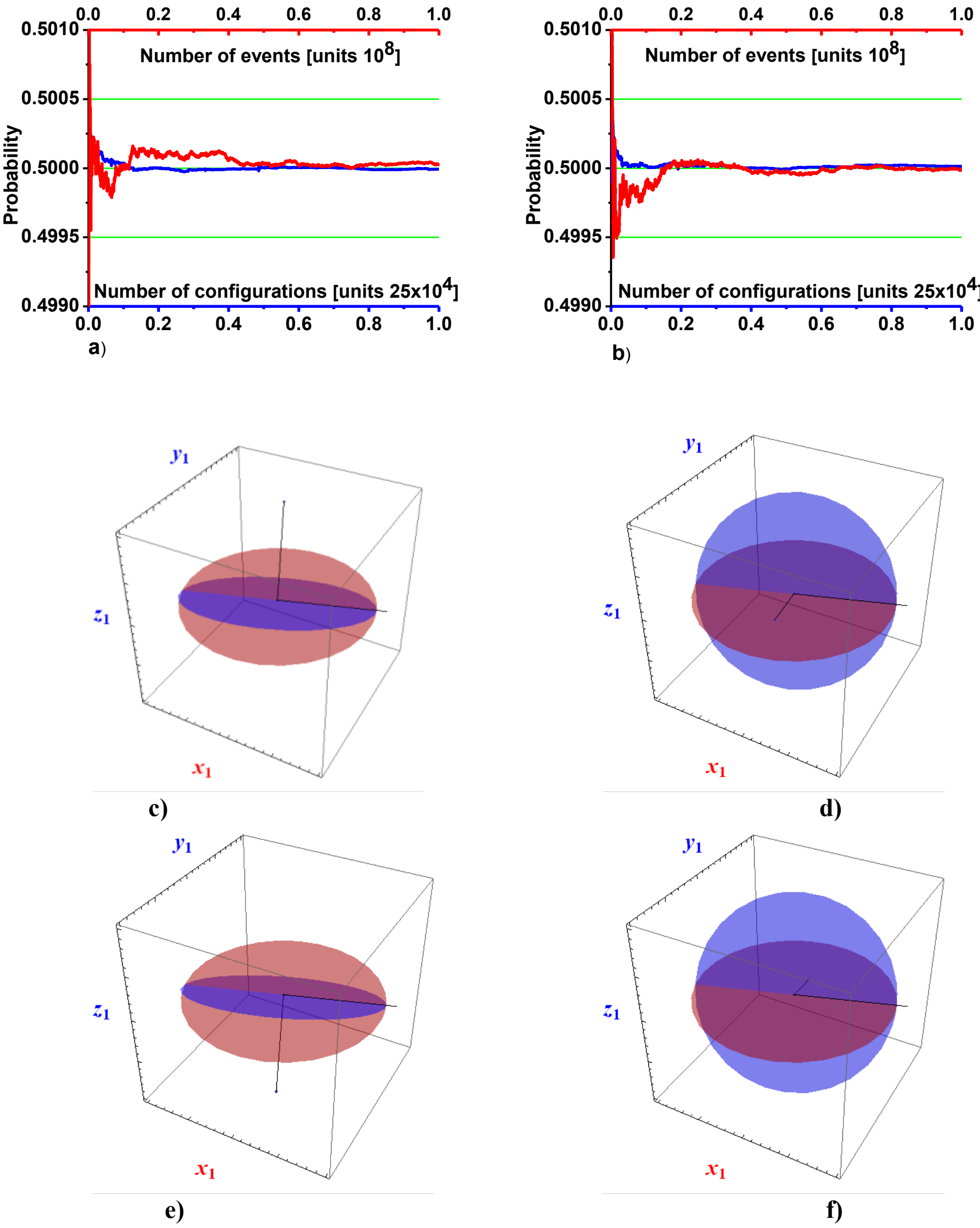

Fig. 3 – a, b – Probability computed by method A (blue) and method B (red) for the case b) ( $\omega_{z0} = 0$ ); c, d, e, f – the initial (red) and at different times $T$ (blue) orientations of the coin. Also, the normal to the coin surface and direction of the angular momentum **S** at different times $T$ are shown. For the parameter values, see the text.

## 4. CONCLUSIONS

We have modeled the coin toss by using a deterministic torque-free rigid body dynamics for constant angular momentum with respect to BF. Using this model, we have simulated the probability that a coin tossed by a usual person ends up the way it started. Under the assumption that there is no random element in the tossing of the coin, we obtain an unbiased probability for the outcome. The numerical code for computing the orientation of the tossed coin and the outcome probability may be asked to the authors.